\newif\ifdraft
\newif\iffull
\newif\ifcomment
\fulltrue
\def\dvers{v1.5}
\documentclass[10pt,aps,prl,twocolumn,superscriptaddress,altaffilletter,nobibnotes,nofootinbib]{revtex4-1}
\usepackage{graphicx}  % needed for figures
\usepackage{dcolumn}   % needed for some tables
\usepackage{bm}        % for math
\usepackage{amssymb}   % for math
\usepackage{amsfonts}
\usepackage{graphics}
\usepackage{epsfig}
\usepackage[usenames]{color}
\newcommand{\ZDC}          {\rm{ZDC}}
\newcommand{\ZDCs}         {\rm{ZDCs}}
\newcommand{\SPD}          {\rm{SPD}}

\newcommand{\VZERO}        {\rm{VZERO}}
\newcommand{\VZEROA}       {\rm{VZERO-A}}
\newcommand{\VZEROC}       {\rm{VZERO-C}}

\newcommand{\pp}           {pp}
\newcommand{\ppbar}        {\mbox{$\mathrm {p\overline{p}}$}}
\newcommand{\PbPb}         {\mbox{Pb--Pb}}
\newcommand{\AuAu}         {\mbox{Au--Au}}

\newcommand{\dNdeta}       {\mathrm{d}N_\mathrm{ch}/\mathrm{d}\eta}
\newcommand{\dNdetatr}     {\mathrm{d}N_\mathrm{tracklets}/\mathrm{d}\eta}

\newcommand{\lum}          {\, \mbox{${\rm cm}^{-2} {\rm s}^{-1}$}}

\newcommand{\pt}           {\ensuremath{p_{\rm t}}}

\newcommand{\snn}          {\ensuremath{\sqrt{s_{\rm NN}}}}
\newcommand{\snnbf}        {\ensuremath{\mathbf{{\sqrt{s_{\mathbf NN}}}}}}

\newcommand{\Npart}        {\ensuremath{N_\mathrm{part}}}
\newcommand{\avNpart}      {\ensuremath{\langle N_\mathrm{part} \rangle}}

\newcommand{\stat}         {({\it stat.})}
\newcommand{\syst}         {({\it sys.})}
\newcommand{\Fig}[1]       {Fig.~\ref{#1}}
\newcommand{\Figure}[1]    {Figure~\ref{#1}}

\newcommand{\warn}[1]      {{\small\textbf{(!\footnote{\textbf{(!)}~#1})}}\marginpar{\textbf{---}}}

\newcommand{\final}[1]     {\textbf{\textcolor{blue}{#1}}}

\graphicspath{{./img/}}
\ifdraft
\usepackage{lineno}
\linenumbers
\modulolinenumbers[5]
\usepackage{fancyhdr}
\renewcommand{\final}[1]{#1}
\else
\renewcommand{\warn}[1]{}
\renewcommand{\final}[1]{#1}
\fi

\usepackage{preprintcover}

\PreprintIdNumber{CERN-PH-EP-2010-060}
\PreprintCoverPaperTitle{Charged-particle multiplicity density at mid-rapidity in central \PbPb\ collisions at \snnbf\ = 2.76 TeV}
\PreprintCoverAbstract{The first measurement of the charged-particle multiplicity density at mid-rapidity
in \PbPb\ collisions at a centre-of-mass energy per nucleon pair \mbox{\snn\ = 2.76 TeV} is presented.
For an event sample corresponding to the most central 5\% of the hadronic cross section
the pseudo-rapidity density of primary charged particles at mid-rapidity is
\final{1584} $\pm$ \final{4} \stat\ $\pm$ \final{76} \syst,
which corresponds to \final{8.3} $\pm$ \final{0.4}~\syst\ per participating nucleon pair.
This represents an increase of about a factor \final{1.9} relative to \pp\ collisions at similar collision 
energies, and about a factor \final{2.2} to central \AuAu\ collisions at $\snn=0.2$~TeV.
This measurement provides the first experimental constraint for models of nucleus--nucleus collisions
at LHC energies.
}
\PreprintJournalName{PRL}

\begin{document}
%==========================================================%
\title{Charged-particle multiplicity density at mid-rapidity \\
in central \PbPb\ collisions at \snnbf\ = 2.76 TeV}
\iffull
% $Id:$
% Author list created by c:\users\kuijer\documents\visual studio 2010\projects\aliautorlist\aliautorlist\alixmldocument.cpp(Nov  9 2010)
% XML file: file:///C:/Documents and Settings/kuijer/Documents/Visual Studio 2010/Projects/AliAutorList/AliAutorList/authors-2010-11-14_flow.xml
% XML created on: 2010-11-14T21:49:13
%
\collaboration{ALICE Collaboration} % CERN-LHC-ALICE
\noaffiliation
\vspace{0.3cm}
\author{K.~Aamodt}
\altaffiliation{}
\affiliation{Department of Physics and Technology, University of Bergen, Bergen, Norway}
\author{B.~Abelev}
\altaffiliation{}
\affiliation{Lawrence Livermore National Laboratory, Livermore, California, United States}
\author{A.~Abrahantes~Quintana}
\altaffiliation{}
\affiliation{Centro de Aplicaciones Tecnol\'{o}gicas y Desarrollo Nuclear (CEADEN), Havana, Cuba}
\author{D.~Adamov\'{a}}
\altaffiliation{}
\affiliation{Nuclear Physics Institute, Academy of Sciences of the Czech Republic, \v{R}e\v{z} u Prahy, Czech Republic}
\author{A.M.~Adare}
\altaffiliation{}
\affiliation{Yale University, New Haven, Connecticut, United States}
\author{M.M.~Aggarwal}
\altaffiliation{}
\affiliation{Physics Department, Panjab University, Chandigarh, India}
\author{G.~Aglieri~Rinella}
\altaffiliation{}
\affiliation{European Organization for Nuclear Research (CERN), Geneva, Switzerland}
\author{A.G.~Agocs}
\altaffiliation{}
\affiliation{KFKI Research Institute for Particle and Nuclear Physics, Hungarian Academy of Sciences, Budapest, Hungary}
\author{S.~Aguilar~Salazar}
\altaffiliation{}
\affiliation{Instituto de F\'{\i}sica, Universidad Nacional Aut\'{o}noma de M\'{e}xico, Mexico City, Mexico}
\author{Z.~Ahammed}
\altaffiliation{}
\affiliation{Variable Energy Cyclotron Centre, Kolkata, India}
% ...
% End of author list

\else
\collaboration{ALICE Collaboration}
\fi
\vspace{0.3cm}
\ifdraft
\date{\today, \color{red}DRAFT \dvers\ \$Revision: 438 $\color{white}:$\$\color{black}}
\else
\date{\today}
\fi
%==========================================================%
\begin{abstract}
The first measurement of the charged-particle multiplicity density at mid-rapidity
in \PbPb\ collisions at a centre-of-mass energy per nucleon pair \mbox{\snn\ = 2.76 TeV} is presented.
For an event sample corresponding to the most central 5\% of the hadronic cross section
the pseudo-rapidity density of primary charged particles at mid-rapidity is
\final{1584} $\pm$ \final{4} \stat\ $\pm$ \final{76} \syst,
which corresponds to \final{8.3} $\pm$ \final{0.4}~\syst\ per participating nucleon pair.
This represents an increase of about a factor \final{1.9} relative to \pp\ collisions at similar collision 
energies, and about a factor \final{2.2} to central \AuAu\ collisions at $\snn=0.2$~TeV.
This measurement provides the first experimental constraint for models of nucleus--nucleus collisions
at LHC energies. %, and enables more precise predictions for a variety of different observables.
\end{abstract}
%\keywords{}
%==========================================================%
\maketitle
\ifdraft
\thispagestyle{fancyplain}
\fi
%==========================================================%
%============================MAIN==========================%
%==========================================================%
% $Id: multMain.tex 329 2010-11-29 20:07:35Z loizides $
\iffull
\newpage
\fi
%%%%%%%%%%%%%%%%%%%%%%%%%%%%%%%%%%%%%%%%%%%%%%%%%%%%%%%%%%%%%%%%%%%%%%%%%%%%%%%%%%%%%%%%%%%%%%%%%%%%
%\section{Introduction}
%%%%%%%%%%%%%%%%%%%%%%%%%%%%%%%%%%%%%%%%%%%%%%%%%%%%%%%%%%%%%%%%%%%%%%%%%%%%%%%%%%%%%%%%%%%%%%%%%%%%
The theory of strong interactions, Quantum Chromo-Dynamics~(QCD), predicts a phase transition
at high temperature between hadronic matter, where quarks and gluons are confined inside hadrons, 
and a deconfined state of matter, the Quark--Gluon Plasma (QGP).
A new frontier in the study of QCD matter opened with the first collisions of $^{208}$Pb ions in November 2010, 
at the Large Hadron Collider~(LHC) at CERN. 
These collisions are expected to generate matter at unprecedented temperatures and energy densities in the laboratory.

The first step in characterizing the system produced in these collisions is the measurement
of the charged-particle pseudo-rapidity density, which constrains the dominant particle production
mechanisms and is essential to estimate the initial energy density.
The dependence of the charged-particle multiplicity density on energy and system size reflects the interplay
between hard parton--parton scattering processes and soft processes.
Predictions of models that successfully describe particle production at RHIC
vary by a factor of two at the LHC~\cite{Abreu:2007kv,Armesto:2009ug}.

This Letter reports the  measurement of the charged-particle pseudo-rapidity density produced in 
\PbPb\ collisions at the LHC, utilizing data taken with the \mbox{ALICE} detector~\cite{aliceapp}
at a centre-of-mass energy per nucleon pair $\snn=2.76$ TeV. 
The primary charged-particle density, $\dNdeta$, in central~(small impact parameter)
\PbPb\ collisions is measured in the pseudo-rapidity interval $|\eta| \equiv |- \ln \tan (\theta/2)| < 0.5$, 
where $\theta$ is the polar angle between the charged-particle direction and the beam axis~($z$).
We define primary particles as prompt particles produced in the collision, including decay products, 
except those from weak decays of strange particles. 

The present measurement extends the study of particle densities in nucleus--nucleus collisions into the TeV regime.
We make comparisons to model predictions~\mbox{\cite{Busza:2007ke,Deng:2010mv,Bopp:2007sa,Mitrovski:2008hb, 
Albacete:2010fs,Levin:2010zy,Kharzeev:2004if,Kharzeev:2007zt,Armesto:2004ud,Eskola:2001bf,Bozek:2010wt,
Sarkisyan:2010kb,Humanic:2010su}}, 
and to previous measurements in nucleus--nucleus collisions at lower energies at the SPS and 
RHIC~\mbox{\cite{Abreu:2002fw, Adler:2001yq, Bearden:2001xw, Bearden:2001qq, Adcox:2000sp, Back:2000gw, 
Back:2001bq, Back:2002wb,Alver:2010ck}}, 
as well as to \pp\ and \ppbar\ collisions over a wide energy 
range~\mbox{\cite{ua1,ua51,ua52,starpp,cdf,alicepp2,Khachatryan:2010xs}}.
Our measurement provides new insight into particle production mechanisms in high energy nuclear collisions 
and enables more precise model predictions for a wide array of other observables in nuclear collisions at the LHC.

%%%%%%%%%%%%%%%%%%%%%%%%%%%%%%%%%%%%%%%%%%%%%%%%%%%%%%%%%%%%%%%%%%%%%%%%%%%%%%%%%%%%%%%%%%%%%%%%%%%%
%\section{ALICE detector and data collection}
%%%%%%%%%%%%%%%%%%%%%%%%%%%%%%%%%%%%%%%%%%%%%%%%%%%%%%%%%%%%%%%%%%%%%%%%%%%%%%%%%%%%%%%%%%%%%%%%%%%%
A detailed description of the ALICE experiment is given in Ref.~\cite{aliceapp}.
Here, we briefly describe the detector components used in this analysis. 
The Silicon Pixel Detector (\SPD) is the innermost part of the Inner Tracking System~(ITS). 
It consists of two cylindrical layers of hybrid silicon pixel assemblies positioned at radial distances 
of $3.9$ and $7.6$~cm from the beam line, with a total of $9.8\times 10^{6}$ pixels of size $50\times425$~$\mu$m$^2$, 
read out by 1200 electronic chips. 
The SPD coverage for particles originating from the center of the detector is $|\eta|<2.0$ and $|\eta|<1.4$
for the inner and outer layers, respectively. 
Each chip provides a fast signal if at least one of its pixels is hit. 
The signals from the 1200 chips are combined in a programmable logic unit which supplies a trigger signal. 
The fraction of SPD channels active during data taking was 70\% for the inner and 78\% for the outer layer. 
The \VZERO\ detector consists of two arrays of 32 scintillator tiles placed at distances $z=3.3$~m and $z=-0.9$~m 
from the nominal interaction point, covering the full azimuth within $2.8<\eta<5.1$~(\VZEROA) and 
$-3.7<\eta<-1.7$~(\VZEROC). 
Both the amplitude and the time signal in each scintillator are recorded.
The \VZERO\ time resolution is better than 1~ns, allowing discrimination of beam--beam collisions 
from background events produced upstream of the experiment. 
The \VZERO\ also provides a trigger signal. 
The Zero Degree Calorimeters (\ZDCs) measure the energy of spectators~(non-interacting nucleons) in two 
identical detectors, located $\pm114$~m from the interaction point. 
Each \ZDC\ consists of two quartz fiber sampling calorimeters: a neutron calorimeter positioned
between the two beam pipes downstream of the first machine dipole that separates the two charged particle beams,
and a proton calorimeter positioned externally to the outgoing beam pipe. 
The energy resolution at beam energy is estimated to be 11\% for the neutron and 13\% for the proton
calorimeter, respectively.

%%%%%%%%%%%%%%%%%%%%%%%%%%%%%%%%%%%%%%%%%%%%%%%%%%%%%%%%%%%%%%%%%%%%%%%%%%%%%%%%%%%%%%%%%%%%%%%%%%%%
%\section{Event selection and centrality}
%%%%%%%%%%%%%%%%%%%%%%%%%%%%%%%%%%%%%%%%%%%%%%%%%%%%%%%%%%%%%%%%%%%%%%%%%%%%%%%%%%%%%%%%%%%%%%%%%%%%
For the data analyzed, beams of four bunches, with about $10^{7}$ Pb ions per bunch, collided at $\snn=2.76$ 
TeV, with an estimated luminosity of $5\times10^{23}$~\lum. 
The trigger was configured for high efficiency for hadronic events, requiring at least two out of the following 
three conditions: 
i) two pixel chips hit in the outer layer of the \SPD, 
ii) a signal in \VZEROA, 
iii) a signal in \VZEROC.
The threshold in the \VZERO\ detector corresponds approximately to the energy deposition of a minimum ionizing 
particle.
The luminous region had an r.m.s.\ width of \final{5.9}~cm in the longitudinal direction 
and \final{50}~$\mu$m in the transverse direction.
The estimated luminosity corresponds to a hadronic collision rate of about \final{4}~Hz. 
The observed rate was about \final{50}~Hz, mainly due to electromagnetically induced processes~\cite{Baur:2001jj}.
These processes have very large cross sections at LHC energies but generate very low multiplicities
and therefore do not contribute to the high particle multiplicities of interest for the present analysis.
The trigger rate without beam was negligible and the rate in coincidence with bunches of only one beam was about
\final{1}~Hz.
This beam background is eliminated from the triggered event sample using the \VZERO\ timing information,
as well as the correlation between the number of tracks reconstructed in the Time Projection Chamber~(TPC)
and the number of hits in the SPD.

\begin{figure}[tbh]
\begin{center}
\includegraphics[width=\linewidth]{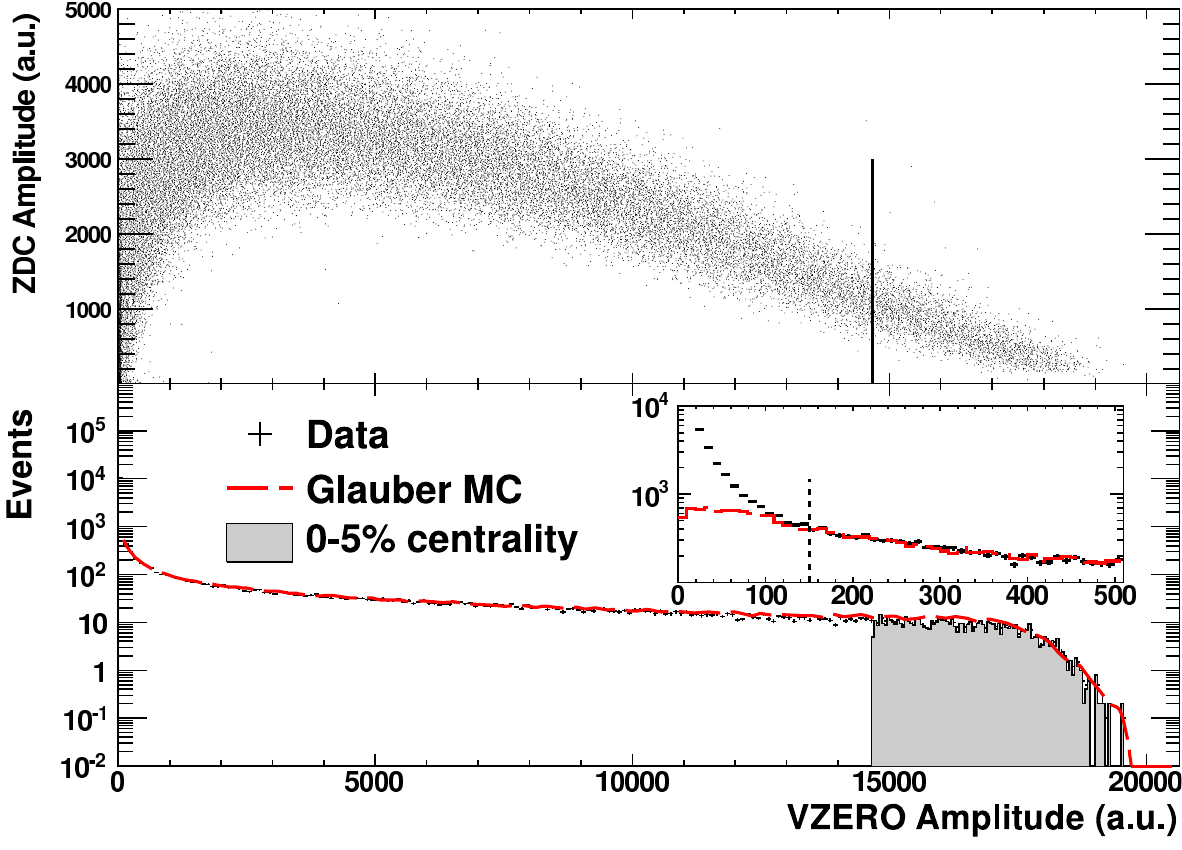}
\caption{\label{fig1}
Upper panel: Correlation of \ZDC\ and the \VZERO\ response in hadronic collisions.
Lower panel: Distribution of the sum of amplitudes in the \VZERO\ scintillator tiles~(black histogram); 
inset shows the low amplitude part of the distribution. 
The red line shows the fit of the Glauber calculation to the measurement.
The fit was performed above the cut indicated in the inset, avoiding the region at low
amplitudes dominated by the electromagnetic processes.
The shaded area corresponds to the most central 5\% of hadronic collisions. 
}
\end{center}
\end{figure}

Offline event characterization utilizes global event observables that are intrinsically correlated 
over different regions of phase space through the initial collision geometry. 
\Figure{fig1}~(upper) shows the measured correlation between the energy deposited in the \ZDC\ 
and the sum of amplitudes in the \VZERO\ detector.  
The \VZERO\ response is proportional to the event multiplicity, and the \ZDC\ energy to the number of
non-interacting nucleons close to beam rapidity.
As events become more central, with smaller impact parameter, they generate larger multiplicity in \VZERO\ 
and less energy forward in the \ZDC. 
This behavior is understood based on collision geometry and nuclear breakup~\cite{Miller:2007ri}. 
For small ZDC response the VZERO signal has two distinct values corresponding to peripheral and central collisions. 
However, the \VZERO\ signal alone can be used to discriminate on centrality.
\Figure{fig1}~(lower) shows the distribution of the \VZERO\ amplitude for all triggered events after 
beam background removal. 
The distribution is fit using the Glauber model~\cite{Alver:2008aq} to describe
the collision geometry and a Negative Binomial Distribution~(NBD) to describe particle production~\cite{Miller:2007ri}.
In addition to the two parameters of the NBD, there is one free parameter that controls the power-law dependence 
of particle production on the number of participating nucleons~($\Npart$).
To avoid the region contaminated by electromagnetic processes, which constitutes over 90\% of the 
triggered events, the fit is restricted to the \VZERO\ amplitude region above \final{150}, where the 
trigger for hadronic collisions is fully efficient.
The fraction of the hadronic cross section from the model fit corresponding to this cut, \final{87}\%, 
allows the determination of the cross section percentile for any more-central \VZERO\ cut by integrating
the measured distribution. 
The most central 5\% fraction of the hadronic cross section was determined in this way.

%%%%%%%%%%%%%%%%%%%%%%%%%%%%%%%%%%%%%%%%%%%%%%%%%%%%%%%%%%%%%%%%%%%%%%%%%%%%%%%%%%%%%%%%%%%%%%%%%%%%
%\section{Data analysis}
%%%%%%%%%%%%%%%%%%%%%%%%%%%%%%%%%%%%%%%%%%%%%%%%%%%%%%%%%%%%%%%%%%%%%%%%%%%%%%%%%%%%%%%%%%%%%%%%%%%%
The analysis is based on the \VZERO\ event selection as described above. 
Among the triggered sample of about 650000 events, \final{3615} events correspond to the most central 5\% 
of the hadronic cross section, indicated by the shaded region in~\Fig{fig1}~(lower).
The first step in the measurement of the charged particle multiplicity is the 
determination of the primary vertex position by correlating hits in the two SPD layers. 
All events in the central sample are found to have a well-constrained primary vertex. 
\begin{figure}[tbh]
\begin{center}
\includegraphics[width=\linewidth]{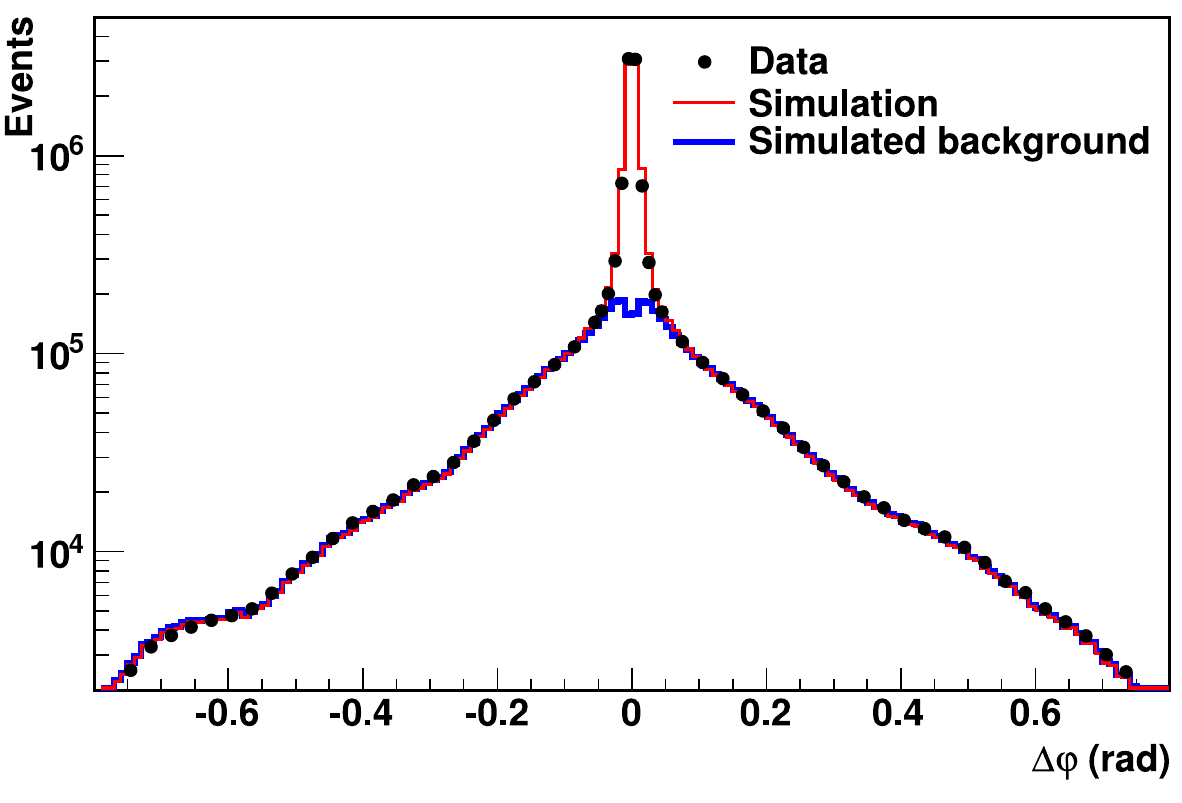}
\caption{\label{fig2} Distribution of the azimuthal separation~($\Delta\varphi$) of all candidate 
tracklets in data, simulation, and the background contribution that is estimated from the simulation.}
\end{center}
\end{figure}
To minimize edge effects at the limit of the SPD acceptance we have only used events with reconstructed 
vertex at $|z_{\rm vtx}|<7$~cm, reducing the sample to \final{2711} events. 
Tracklet candidates~\cite{Alver:2010ck} are formed using information on the position of the primary vertex 
and of hits on the SPD layers. 
A tracklet is defined by a pair of hits, one on each SPD layer.
Using the reconstructed vertex as the origin, we calculate the differences in azimuthal~($\Delta\varphi$, 
bending plane) and polar~($\Delta\theta$, non-bending direction) angles for pairs of hits~\cite{alicepp2}. 
Only hit combinations satisfying a selection on the sum of the squares of $\Delta\varphi$ and $\Delta\theta$, 
each normalized to its estimated resolution~(60~mrad in $\Delta\varphi$ and $25\,\rm{sin}^{2}\theta$~mrad in 
$\Delta\theta$), are selected as tracklets.
If multiple tracklet candidates share a hit, only the combination with the smallest sum of squares
of $\Delta\varphi$ and $\Delta\theta$ is kept. 
The cut imposed on $\Delta\varphi$ efficiently selects charged particles with transverse momentum~(\pt) 
above 50~MeV/$c$. 
Particles below 50~MeV/$c$ are mostly absorbed by material. 

The charged-particle pseudo-rapidity density $\dNdeta$ is obtained from the number of tracklets 
within $|\eta|<0.5$ according to $\dNdeta=\alpha\times(1-\beta)\dNdetatr$, 
where $\alpha$ is the correction factor for the acceptance and efficiency for a primary track 
to generate a tracklet and $\beta$ is the probability to form a background tracklet from uncorrelated hits. 
The corrections~$\alpha$ and $\beta$ are determined as a function of the $z$-position of the primary vertex 
and the pseudo-rapidity of the tracklet.
The simulations used to calculate the corrections are based on the HIJING~\cite{hijing} event generator
and a GEANT3~\cite{geant3ref2} model of the detector response.
Three different methods have been used to estimate the combinatorial background.

The main method to estimate the combinatorial background $\beta$ relies on the event simulation 
using a sample of events with similar multiplicities~(SPD hits) as in the real data. 
In \Fig{fig2} the $\Delta\varphi$ distribution for candidate tracklets is compared for data and 
simulation.
The distributions are very similar, practically identical in the background dominated tails.
The second method is based on the injection of random background hits in the real event, 
in order to evaluate the probability of creating fake tracklets by combinatorics. 
%In the second method, this is achieved by combining hits taking different events for the inner and 
%outer SPD layers~(chosen with similar values of the vertex position and multiplicity).
\ifcomment
\begin{table}[tbh!f]
\centering{
\begin{tabular}{lc}
\hline
Source                       & Uncertainty\\
\hline
%Tracklet selection cuts      & negl.\\
%Material budget              & negl.\\
%Misalignment                 & negl.\\
Background subtraction       & \final{2}\%\\
Particle composition         & \final{1}\%\\
Contamination by weak decays & \final{1}\%\\
Low-$\pt$ extrapolation      & \final{2}\%\\
%Detector efficiency          & negl.\\
Event generator		     & \final{2}\%\\
Centrality definition        & \final{3}\%\\
%Background events            & negl.\\
%Triggering efficiency        & negl.\\
\hline\hline
Total~(added in quadrature)  & \final{4.8}\%\\
\hline
\end{tabular}
\caption{\label{systable}Contributions to the systematic uncertainty on the charged-particle 
pseudo-rapidity density in $|\eta|<0.5$.}}
\end{table}
\fi
In the third method events are modified by rotating hits in the inner SPD layer by 180$^\mathrm{o}$ in $\varphi$, 
thereby destroying real correlations, but preserving global event features.
In all cases, the absolute amount of combinatorial background is obtained by matching the tracklet and background 
distributions in the tails. 
For the main method, which ideally provides both the shape and the normalization, an adjustment of \final{1}\% 
is needed to match the tails.
The estimated combinatorial background is about \final{14}\%.
In order to account for the effect of the correlated background, the same background subtraction procedure is
also applied to the simulation~(i.e.\ without relying on the event generator information).
The correction for the acceptance and efficiency, $\alpha$, is obtained by the ratio of the number of generated 
primary charged particles to the number of reconstructed tracklets after subtraction of the combinatorial background.
In this way, $\alpha$ accounts for geometrical acceptance, detector and reconstruction efficiencies,
contamination by weak decay products of strange particles, conversions, secondary interactions
and undetected particles below 50 MeV/$c$ transverse momentum.
The overall correction factor $\alpha$ varies slightly depending on vertex position and $\eta$,
and is about \final{2}.

We have considered the following sources of systematic uncertainties: %~(Table~\ref{systable}):
background subtraction estimated as 2\% by comparing the results of different methods;
particle composition estimated as 1\% by changing the relative abundances of protons, pions, kaons 
by a factor of two;
contamination by weak decays estimated as 1\% by changing the relative contribution of the yield of strange particles
by a factor of two; 
low-$\pt$ extrapolation estimated as 2\% by varying the amount of undetected particles at low $\pt$ by a factor of two;
event generator estimated as 2\% by using HIJING~\cite{hijing} with and without jet quenching, 
as well as \mbox{DPMJET}~\cite{Roesler:2000he} for the corrections;
centrality definition estimated as 3\% by using an alternative event selection based on the SPD hit multiplicities,
and by varying the range of the Glauber model fit. %Matching with HIJING is found to be compatible.
All other sources of systematic errors considered~(tracklet cuts, vertex cuts, material budget, detector
efficiency, background events) were found to be negligible.
The total systematic errors amounts to $4.8$\%.
Independent cross-checks performed using tracks reconstructed in the TPC and ITS
yield results consistent within the systematic uncertainty.

%%%%%%%%%%%%%%%%%%%%%%%%%%%%%%%%%%%%%%%%%%%%%%%%%%%%%%%%%%%%%%%%%%%%%%%%%%%%%%%%%%%%%%%%%%%%%%%%%%%%
%\section{Results}
%%%%%%%%%%%%%%%%%%%%%%%%%%%%%%%%%%%%%%%%%%%%%%%%%%%%%%%%%%%%%%%%%%%%%%%%%%%%%%%%%%%%%%%%%%%%%%%%%%%%
\begin{figure}[tbh!]
\begin{center}
\includegraphics[width=\linewidth]{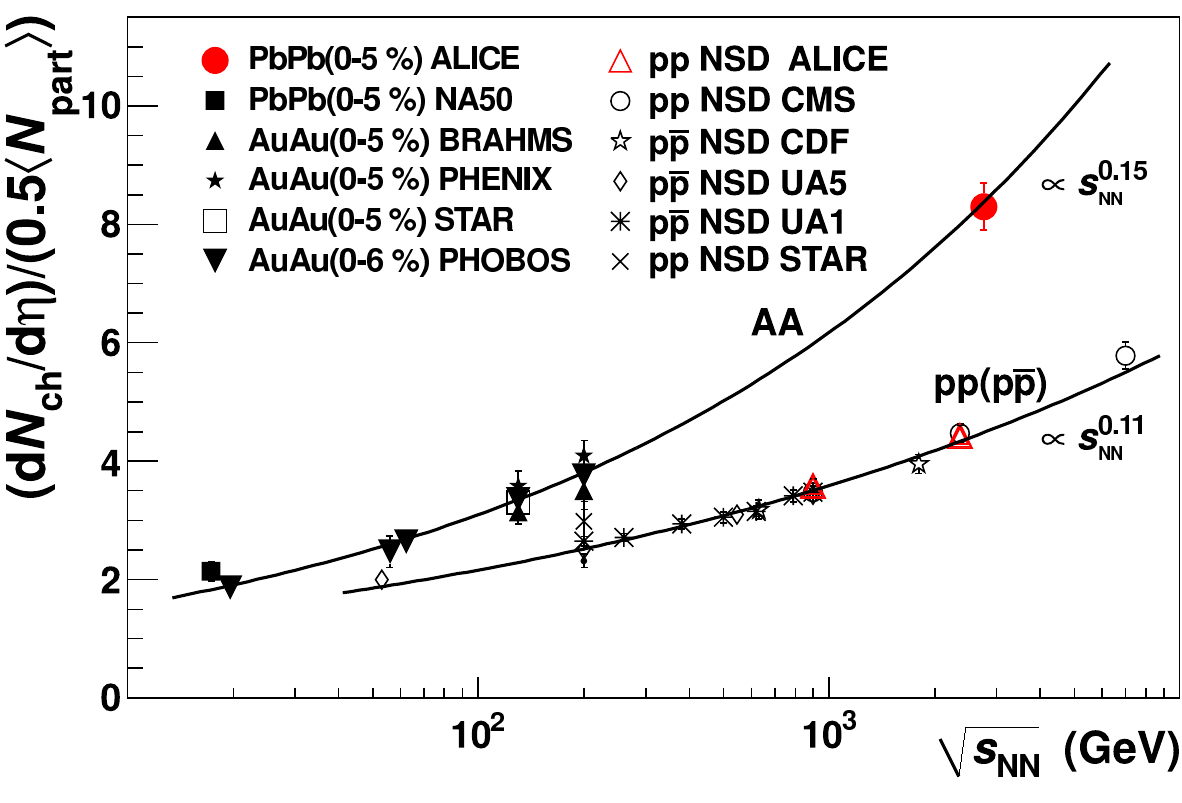}
\caption{\label{fig3}
Charged particle pseudo-rapidity density per participant pair for central nucleus--nucleus~\cite{Abreu:2002fw, 
Adler:2001yq,Bearden:2001xw,Bearden:2001qq,Adcox:2000sp,Back:2000gw,Back:2001bq,Back:2002wb,Alver:2010ck}
and non-single diffractive \pp~(\ppbar) collisions~\cite{ua1,ua51,ua52,starpp,cdf,alicepp2,Khachatryan:2010xs}, 
as a function of $\snn$.
The solid lines $\propto s_{\rm NN}^{0.15}$ and $\propto s_{\rm NN}^{0.11}$ are superimposed on the
heavy-ion and \pp~(\ppbar) data, respectively.}
\end{center}
\end{figure}
In order to compare bulk particle production in different collision systems and at different energies, 
and to compare with model calculations, the charged particle density is scaled by the number 
of participating nucleons, determined using the Glauber model fit described above~(\Fig{fig1}).
The average number of participants for the 5\% most central events is found to be
$\avNpart = $ \final{381} with an r.m.s.\ of 18 and a systematic uncertainty of \final{1}\%.
The systematic uncertainty was obtained by varying the parameters of the Glauber calculation
within the experimental uncertainty and by $\pm8$\% around $64$~mb for the nucleon--nucleon cross section,
by using different fit ranges, and by comparing results obtained for different centrality variables~(SPD 
hits, or combined use of the \ZDC\ and \VZERO\ signals).

We measure a density of primary charged particles at mid-rapidity
$\dNdeta=$ \final{1584} $\pm$ \final{4} \stat\ $\pm$ \final{76} \syst.
Normalizing per participant pair, we obtain
$\dNdeta/(0.5\,\avNpart)= $ \final{8.3} $\pm$ \final{0.4}~\syst\ with negligible statistical error.
In \Fig{fig3}, this value is compared to the measurements for \AuAu~and \PbPb, and 
non-single diffractive~(NSD) \pp\ and \ppbar\ collisions over a wide range of 
collision energies~\cite{Abreu:2002fw,Adler:2001yq,Bearden:2001xw, 
Bearden:2001qq,Adcox:2000sp,Back:2000gw,Back:2001bq,Back:2002wb,Alver:2010ck,
ua1,ua51,ua52,starpp,cdf,alicepp2,Khachatryan:2010xs}.
It is interesting to note that the energy dependence is steeper for heavy-ion collisions
than for \pp\ and \ppbar\ collisions.
For illustration, the curves $\propto s_{\rm NN}^{0.15}$ and $\propto s_{\rm NN}^{0.11}$ are shown
superimposed on the data.
A significant increase, by a factor \final{2.2}, in the pseudo-rapidity density is observed at 
$\snn = 2.76$ TeV for \PbPb\ compared to $\snn=0.2$~TeV for \AuAu.
The average multiplicity per participant pair for our centrality selection is found to be a factor \final{1.9} 
higher than that for \pp\ and \ppbar\ collisions at similar energies.

\begin{figure}[tbh]
\begin{center}
\includegraphics[width=\linewidth]{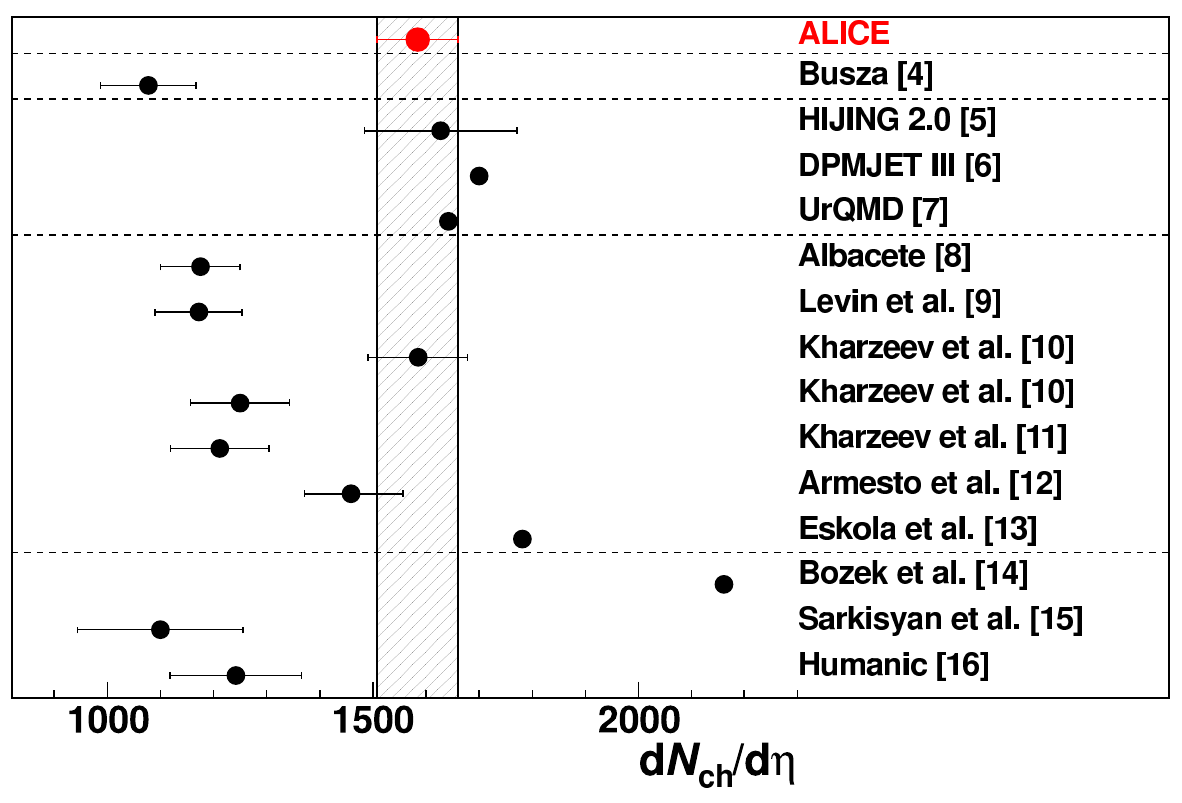}
\caption{\label{fig4} Comparison of this measurement with model predictions. 
Dashed lines group similar theoretical approaches.}
\end{center}
\end{figure}

\Figure{fig4} compares the measured pseudo-rapidity density to model calculations that describe RHIC measurements
at $\snn=0.2$~TeV, and for which predictions at $\snn=2.76$~TeV are available.
Empirical extrapolation from lower energy data~\cite{Busza:2007ke} significantly underpredicts the measurement.
Perturbative QCD-inspired Monte Carlo event generators, based on the HIJING model tuned to 7 TeV \pp\ data without
jet quenching~\cite{Deng:2010mv}, on the Dual Parton Model~\cite{Bopp:2007sa}, or on the Ultrarelativistic 
Quantum Molecular Dynamics model~\cite{Mitrovski:2008hb} are consistent with the measurement.
Models based on initial-state gluon density saturation have a range of predictions depending on the specific 
implementation~\cite{Albacete:2010fs,Levin:2010zy,Kharzeev:2004if,Kharzeev:2007zt,Armesto:2004ud},  
and exhibit a varying level of agreement with the measurement. 
The prediction of a hybrid model based on hydrodynamics and saturation of final-state phase space of scattered 
partons~\cite{Eskola:2001bf} is close to the measurement. 
A hydrodynamic model in which multiplicity is scaled from p+p collisions overpredicts the 
measurement~\cite{Bozek:2010wt}, while a model incorporating scaling based on Landau hydrodynamics 
underpredicts the measurement~\cite{Sarkisyan:2010kb}. Finally, a calculation based on modified PYTHIA and 
hadronic rescattering~\cite{Humanic:2010su} underpredicts the measurement.

In summary, we have measured the charged-particle pseudo-rapidity density at mid-rapidity in \PbPb\ collisions 
at $\snn = 2.76$~TeV, for the most central 5\% fraction of the hadronic cross section. 
We find $\dNdeta = $ \final{1584} $\pm$ \final{4}~\stat\ $\pm$ \final{76}~\syst, corresponding to 
\final{8.3} $\pm$ \final{0.4}~\syst\ per participant pair.
These values are significantly larger than those measured at RHIC, and indicate a stronger energy dependence 
than measured in \pp\ collisions.  
The result presented in this Letter provides an essential constraint for models describing high energy 
nucleus--nucleus collisions.

%==========================================================%
%==================ACKNOWLEDGEMENTS========================%
%==========================================================%
\iffull
% $Id:$

The ALICE collaboration would like to thank all its engineers and technicians for their invaluable contributions to the construction of the experiment and the CERN accelerator teams for the outstanding performance of the LHC complex.
The ALICE collaboration acknowledges the following funding agencies for their support in building and
running the ALICE detector:
Calouste Gulbenkian Foundation from Lisbon and Swiss Fonds Kidagan, Armenia;
Conselho Nacional de Desenvolvimento Cient\'{\i}fico e Tecnol\'{o}gico (CNPq), Financiadora de Estudos e Projetos (FINEP),
Funda\c{c}\~{a}o de Amparo \`{a} Pesquisa do Estado de S\~{a}o Paulo (FAPESP);
National Natural Science Foundation of China (NSFC), the Chinese Ministry of Education (CMOE)
and the Ministry of Science and Technology of China (MSTC);
Ministry of Education and Youth of the Czech Republic;
Danish Natural Science Research Council, the Carlsberg Foundation and the Danish National Research Foundation;
The European Research Council under the European Community's Seventh Framework Programme;
Helsinki Institute of Physics and the Academy of Finland;
French CNRS-IN2P3, the `Region Pays de Loire', `Region Alsace', `Region Auvergne' and CEA, France;
German BMBF and the Helmholtz Association;
Hungarian OTKA and National Office for Research and Technology (NKTH);
Department of Atomic Energy and Department of Science and Technology of the Government of India;
Istituto Nazionale di Fisica Nucleare (INFN) of Italy;
MEXT Grant-in-Aid for Specially Promoted Research, Ja\-pan;
Joint Institute for Nuclear Research, Dubna;
 %
%Korea Foundation for International Cooperation of Science and Technology (KICOS);
National Research Foundation of Korea (NRF);
CONACYT, DGAPA, M\'{e}xico, ALFA-EC and the HELEN Program (High-Energy physics Latin-American--European Network);
Stichting voor Fundamenteel Onderzoek der Materie (FOM) and the Nederlandse Organisatie voor Wetenschappelijk Onderzoek (NWO), Netherlands;
Research Council of Norway (NFR);
Polish Ministry of Science and Higher Education;
National Authority for Scientific Research - NASR (Autoritatea Na\c{t}ional\u{a} pentru Cercetare \c{S}tiin\c{t}ific\u{a} - ANCS);
Federal Agency of Science of the Ministry of Education and Science of Russian Federation, International Science and
Technology Center, Russian Academy of Sciences, Russian Federal Agency of Atomic Energy, Russian Federal Agency for Science and Innovations and CERN-INTAS;
Ministry of Education of Slovakia;
CIEMAT, EELA, Ministerio de Educaci\'{o}n y Ciencia of Spain, Xunta de Galicia (Conseller\'{\i}a de Educaci\'{o}n),
CEA\-DEN, Cubaenerg\'{\i}a, Cuba, and IAEA (International Atomic Energy Agency);
The Ministry of Science and Technology and the National Research Foundation (NRF), South Africa;
Swedish Reseach Council (VR) and Knut $\&$ Alice Wallenberg Foundation (KAW);
Ukraine Ministry of Education and Science;
United Kingdom Science and Technology Facilities Council (STFC);
The United States Department of Energy, the United States National
Science Foundation, the State of Texas, and the State of Ohio.
\fi
%==========================================================%
%==================BIBLIOGRAPHY============================%
%==========================================================%
\iffull

\else
\bibliographystyle{apsrev4-1}
\bibliography{multPbPb}{}
\fi
%==========================================================%
\end{document}